# Metadata aggregation and "automated digital libraries": A retrospective on the NSDL experience

Carl Lagoze
Cornell Information Science
301 College Ave.
Ithaca, NY 14850
+1-607-255-6046

lagoze@cs.cornell.edu

Dean Krafft, Tim Cornwell,
Naomi Dushay, Dean Eckstrom
Cornell Information Science
301 College Ave.
Ithaca, NY 14850
+1-607-255-5925

{dean,cornwell,naomi,eckstrom}
@cs.cornell.edu

John Saylor
Cornell University Library.
Ithaca, NY 14853
+1-607-255-4134

jms1@cornell.edu

## ABSTRACT
Over three years ago, the Core Integration team of the National Science Digital Library (NSDL) implemented a digital library based on metadata aggregation using Dublin Core and OAI-PMH. The initial expectation was that such low-barrier technologies would be relatively easy to automate and administer. While this architectural choice permitted rapid deployment of a production NSDL, our three years of experience have contradicted our original expectations of easy automation and low people cost. We have learned that alleged "low-barrier" standards are often harder to deploy than expected. In this paper we report on this experience and comment on the general cost, the functionality, and the ultimate effectiveness of this architecture.

## Categories and Subject Descriptors
H.3.7 [**Information Storage and Retrieval**]: Digital Libraries – *collection, dissemination, standards, systems issues.*

## 1. INTRODUCTION
Over the past four years, the NSDL Core Integration team (CI) has developed and administered an expanding education-focused digital library. The visible presence of this digital library is the main NSDL portal[1]. Underlying this portal is an architecture based on the aggregation of metadata from multiple sources, the storage of that metadata in a metadata repository (MR), and the provision of services that consume and process that metadata. One of these services is a Lucene-based search engine that indexes metadata in the MR and, if possible, the full-text content that the metadata references. The NSDL architecture was initially described in an earlier paper [18].

Our choice of this architecture was motivated by a mixture of factors:

*Expediency:* The NSF grant to CI mandated the launch of a production NSDL presence soon after the initiation of funding. This required that the system use established tools and standards, and that it embody familiar practices rather than innovative techniques. We adopted OAI-PMH[2] and Dublin Core[3] based on these criteria. Similarly, we implemented the MR in an Oracle® RDBMS because it permitted the use of familiar "enterprise" system management techniques. Finally, because the metadata-based architecture resembled the well-exercised union cataloging model, we believed that production methods from that model could be used in the NSDL environment. We recognized that these methods would have to be modified due to the differences in complexity between Dublin Core records and library cataloging records and because metadata creators in the NSDL were both widely distributed and were generally not professional catalogers. These design choices were successful in meeting the rapid deployment mandate – the "initial launch" of the NSDL occurred in December 2002, a little over a year after the initiation of CI funding.

*Philosophy:* The choice of structured metadata and the union catalog paradigm reflected principles within the CI team. From the beginning we intended that the initial architecture would evolve to a "spectrum of interoperability" [4], which would accommodate other less traditional paradigms (e.g., focused crawling, automated classification). However, many members of the CI team felt that structured metadata should be at the core of a production digital library. Like many mainstream digital library efforts, they had confidence that structured metadata was a well-known and easily exploited means of making precise information available to library services, such as search and discovery.

*Finances:* Finally, the initial decisions about how to build the NSDL reflected the nature of the CI budget. Over the years, CI has received approximately 4M USD annually from the NSF, with the expectation that most of this would be used for library development, rather than day-to-day operations. This mandated an operational strategy that relied on automation, exploiting relatively inexpensive computers and networks, rather than on expensive human effort [2]. In general, cataloging has been a human-intensive activity in libraries [12]. "Low-barrier" standards such as Dublin Core and OAI-PMH were designed to reduce and distribute this cost, and the NSDL built a library based on such expectations.

---

[1] http://nsdl.org

[2] http://www.openarchives.org

[3] http://dublincore.org

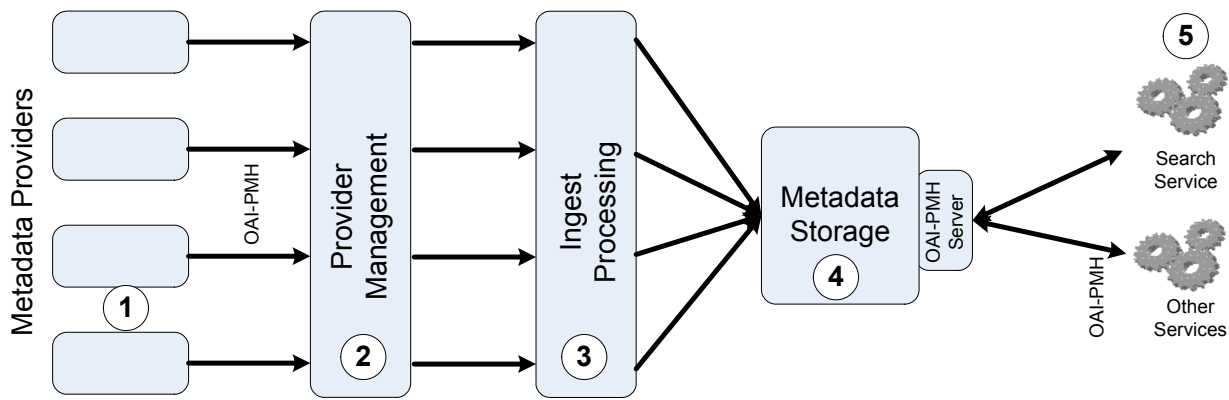

**Figure 1 - Metadata Flow**

As noted, the availability and relative simplicity of the individual architectural components facilitated rapid deployment. However, our three years of experience with the NSDL have contradicted our original expectations of automation and low people cost. We have learned that "low-barrier" standards have been more difficult for contributors to use than expected. Moreover, despite the relatively simplicity of the individual components in the NSDL, the combination of these components, plus maintaining them on a 24x7 basis, adds up to a system of surprising complexity. There are multiple data feeds, many software components, and multiple machines that are distributed over multiple organizations and locations. The number of components and variables to be managed has frequently interfered with our efforts to handle the process automatically, forcing us to fall back on "expensive" human intervention. At times, this human effort has consumed developer time that otherwise could have been used to widen the spectrum of interoperability and innovation.

This paper provides a retrospective on these three years of running a relatively large-scale digital library (over a million objects) by collecting, processing, storing, and using metadata. Our intent is not to argue for or against the utility of metadata aggregation as the basis for a digital library. Such an argument needs to take into account metrics on how metadata actually improves services such as information retrieval (in the manner of the seminal Cranfield experiments [7, 8]), and contrast the costs and benefits of metadata aggregation against other approaches. What we do provide is quantitative and anecdotal data on the operational costs of a metadata-based digital library. Both costs and benefits need to be accounted for in a final evaluation of the architecture.

The organization of this paper reflects the stages in the flow of metadata through the NSDL architecture. It examines each stage, from original provision of metadata to the use of the metadata by services, exemplified by the search service. The description of each phase describes impediments encountered and the success and/or failure of various tools to overcome those impediments. We purposely omit a discussion of user interface portals, since evaluation of user interfaces is by nature different from the system issues that are the focus here.

The metadata flow is shown in Figure 1. The components of this flow, which are labeled with circled numbers the figure, and which correspond to remaining sections in this paper, are as follows.

*(1) Metadata Providers* – These are the organizations from which CI harvests metadata via OAI-PMH. In some cases these metadata providers also manage the content described by the metadata; in others, they either exclusively or additionally aggregate metadata about resources managed by other organizations. The NSDL architecture does not distinguish among these roles; everyone is treated as a metadata provider. We describe problems that providers have encountered with metadata and OAI-PMH, and some tools and techniques that simplified that process.

*(2) Provider Management* – CI has developed a software component known as the Collection Registration Service (CRS) that maintains knowledge of all data providers, descriptions of their collections, their OAI servers and harvest information, harvest schedules, and logs of harvests performed. The intent of this system is to automate, as much as possible, periodic harvests from a large number (potentially hundreds) of OAI-PMH data providers. We will describe issues that arose that have interfered with this automation.

*(3) Ingest Processing* – CI developed back-end services to process the raw OAI-PMH feeds and normalize the metadata before storage in the RDBMS metadata repository. We describe these processes and their efficacy in automating the OAI feeds and improving the metadata.

*(4) Metadata Storage and OAI Re-Exposure* –We describe some aspects of table design of the Oracle-based metadata repository, especially related to the exposure of the metadata via OAI-

PMH[4]. allowing CI to act as an "OAI-PMH aggregator" [21], effectively combining the roles of data provider and service provider. We describe our experiences running a relatively large-scale (1.2 million metadata records) OAI-PMH server and our techniques for linking that server to an RDBMS.

*(5) Search Service*[5] – CI runs a search service that uses Lucene[6] for indexing and query processing. Lucene indexes both the metadata, consumed from the MR via OAI-PMH, and if possible the full-text resources, crawled via Nutch[7] using resource URLs provided in metadata records (if present and accessible). By and large, users of the NSDL (and most libraries) are concerned with finding and accessing resources. As such, the search service (and many other services) needs to translate the metadata-centric data model (where metadata originates from both content holders and metadata aggregators) to a resource-centric view. We will describe issues related to presenting a resource-centric view of the library over a metadata-centric architecture.

The paper concludes with some broader comments on the overall utility of this digital library architecture. Our recent work in the NSDL and other projects [19, 20] focuses on a resource-centric architecture that integrates less structured forms of information, which collectively add value and context to digital library resources. Traditional structured metadata plays a role in such information contextualization. However, it exists as a component of a resource-centric model, rather than being the focus of the information model itself.

## 2. RELATED WORK

The architecture of the NSDL and the issues of metadata creation, harvesting, and aggregation have been described in earlier papers by the CI team. The initial prototype of the architecture was described in [4]. The current NSDL production architecture was introduced in [18]. Some of the processes described in this paper and related issues with metadata aggregation in the NSDL were described in [3]. This paper logically follows after those papers, providing an overview of the costs, problems, and experiences in supporting the metadata aggregation model over the past three years. It also is written at a time when the CI team is engaged in a major project to shift the architecture to a different, resource-centric, paradigm [19]. As such, it provides the opportunity to look back on the initial architecture from the perspective of lessons learned.

The issue of metadata quality is an important factor in the system described here. Even if all other aspects of the system worked perfectly, poor quality metadata would degrade the quality of the resulting library. Diane Hillmann, who was instrumental in the deployment of the NSDL, has written extensively on this issue [6, 17]. With Naomi Dushay, she has written about visualization tools for analyzing the quality of metadata [10]. Other papers focus on the quality of metadata harvested and federated from distributed sources [9, 25]. This paper does not cover metadata quality per se, but touches on it as one of the system design issues, complexities, and costs in maintaining a relatively large-scale metadata aggregation site.

Finally, OAI-PMH, upon which the NSDL system is built, is a de facto standard for metadata sharing about which much has been written. The OAIster system [14] is another example of a large-scale aggregation system. [15] reports findings on a number of metadata harvesting experiments. There has been some research related to normalizing and enhancing large-scale harvests. [13] describes the use of harvested collection metadata records to enhance harvested item records. [16] provides preliminary findings on eliminating duplicates in harvested OAI-PMH records. This paper briefly touches on these issues, but does not focus on them.

## 3. METADATA PROVIDERS

According to the NSDL Collection Development Policy [24], the "NSDL Collection is a *collection* of *sets of resources*. These sets of resources are also referred to as *collections*." Furthermore: "As a general rule, collections that are considered to be part of the NSDL Collection are not actually held within NSDL-owned computers or storage systems. Instead, individual collections typically are held and managed by their owners or providers."

In lieu of storing the resources that make up the NSDL collection, the decision was made to develop and manage a repository of metadata surrogates for these resources. Intentionally operating without a cataloging staff, CI assumed that metadata records would be contributed by external parties, both those that wanted to contribute their content to the NSDL collection, and those that had metadata about other organizations' resources.

The practice of collecting resource surrogates from distributed parties and cataloging them is well established in the library community. OCLC's WorldCat[8] collects and distributes many library catalog records. Our plan was to adapt this model with Dublin Core as a minimalist metadata format, which could be supplemented by richer metadata formats, and OAI-PMH as a low-barrier transport technology. Our expectation was that Dublin Core and OAI-PMH were relatively simple and that surely every collection provider would be able to implement them and be integrated into the NSDL.

In fact, reality fell far short of our expectations. We discovered that the WorldCat paradigm, which works so well in the shared professional culture of the library, is less effective in a context of widely varied commitment and expertise. Few collections were willing or able to allocate sufficient human resources to provide quality metadata. A mandate from the NSF in 2004 that NSF-funded NSDL collections had to share metadata addressed some of the "willingness" problems of those collections. Unfortunately, commercial providers of STEM resources were especially resistant to sharing their metadata; they had yet to learn (e.g. from Google) that open access to discovery information leads to more use (i.e. sales).

---

[4] The baseURL of the NSDL OAI server is http://services.nsdl.org:8080/nsdloai/OAI.

[5] Although there are other services in the NSDL, such as an archive service, we will not describe them in this paper.

[6] http://lucene.apache.org/

[7] http://lucene.apache.org/nutch/

[8] http://www.oclc.org/worldcat/

But more problematic was the reality that the personnel requirements to share metadata were deceptively high due to what can be characterized as a "knowledge gap". Successful provision of metadata actually involves three distinct skill sets:

1. *Domain expertise* – knowledge of the resources themselves and their pedagogical goal.
2. *Metadata expertise* – knowledge of cataloging practices such as use of controlled vocabularies and proper formatting of data such as names and dates.
3. *Technical expertise* – knowledge of tools involved in setting up and running an OAI-PMH server including XML, XML schema, UTF8, and HTTP.

We found that very few NSDL collections had a single person, let alone a team, with these three skill sets. In fact, the "team" for many collections consisted of one person working part-time. Thus, the CI team, which indeed had the combined expertise, had to provide intensive consultation. Documentation on Dublin Core [11] and OAI-PMH [1] helped somewhat, but still the amount of hand-holding was well beyond what was anticipated. An analysis of our collection development email logs indicates that for a large number of collections the time lag between first contact and successful provision of metadata *exceeded several months*, and in one exceptional case spanned two years! Throughout this interim, the CI team had to engage in frequent training and persuasion to move metadata providers into the production cycle.

Some of the technical barriers were overcome by funding from NSF for the development and deployment of the Collection Workflow Integration System (CWIS)[9] "… software to assemble, organize, and share collections of data about resources, like Yahoo! or Google Directory but conforming to international and academic standards for metadata." The CWIS software comes complete with an OAI-PMH server, so that metadata stored within a CWIS installation could be readily ingested into the MR (or any other OAI-PMH aggregator). CWIS has proven effective for some collections and has been deployed on a relatively modest basis. At last count, sixteen NSDL collections, out of the approximately 85 OAI servers, are running CWIS.

Obviously massively scaled web search engines such as Google and Yahoo do not incur either the resistance or costs of metadata provision and harvesting. Although there are limits to automated crawling and indexing – e.g., deep web invisibility and indexing non-textual resources[10] - we recognize that the future of collection development in the NSDL relies on deploying these technologies as a supplement and, in many cases, a replacement for the harvesting model. We are currently working with the iVia project [23] at the University of California-Riverside, which has developed technology for focused crawling, automated metadata generation, and "rich-text" generation (intended for resource discovery). CI has started to use this tool for collection building.

---

[9] http://scout.wisc.edu/Projects/CWIS/

[10] We note that these are not insurmountable limitations and future considerations about metadata and metadata harvesting must consider rapid technical advancements in these areas.

## 4. PROVIDER MANAGEMENT

From a technical perspective, an NSDL *collection* is an entity from which metadata is harvested via OAI-PMH. The Collection Registration Service (CRS) provides a set of services for identifying and managing these collections and for managing the processes that harvest metadata from them. The CRS accomplishes this by maintaining both Dublin Core metadata about the collection itself and the information needed to automatically harvest OAI metadata from the collection's OAI provider.

In this section we describe the design of the automated harvesting system, enumerate some problems with automation, and then describe some statistics related to our harvesting experience.

### 4.1 Automated harvesting model

In the original model, harvesting of metadata was intended to be almost completely automated, with the following workflow:

1. New collections validate their OAI-PMH server[11].
2. A metadata record describing the collection is created and stored in the CRS, which then ports it to the MR.
3. A metadata harvesting record for the collection is created that lists the OAI source, OAI set and format information, provider emails, and a harvest schedule. This record is the basis for automated harvesting
4. An initial full harvest of the collection is initiated.
5. Subsequently, incremental harvests happen on a schedule appropriate to the collection (e.g. weekly, monthly, quarterly), with automatic emails to the provider describing any problems encountered, allowing the provider to correct the problem and schedule an updated harvest.

### 4.2 Automation problems

In a few cases this workflow proceeds smoothly, but the vast majority of cases require significant manual intervention. A detailed enumeration of these problems is impossible due to the constraints of this paper, but we highlight the following.

The process of initially validating a new OAI provider is extensive, typically requiring several email exchanges and repeated harvest attempts. Validation errors run the gamut, including UTF-8 errors, XML schema validation problems, URL and XML encoding problems, improper date stamping, bad resumption tokens, and the like. We provide more details on validation statistics in [26]. Compared to other protocols OAI-PMH may be "low barrier", but deploying it requires reasonable technical sophistication with protocols, XML, schema, and the like.

Providers often fail to use available OAI validation tools, and rarely perform routine self-validation. This places the burden on harvesters like CI to notify providers of problems.

Often validation of an OAI server will fail over time. Because OAI-PMH responses are structured as a set of packages (e.g., "about" containers, metadata) that are variable across OAI-PMH transactions, validation may break down as the content of a package varies, or due to web server upgrades or other software changes.

---

[11] The NSDL wrote its own OAI validator (publicly available at http://repository.comm.nsdl.org/prs_web/harvest_server_val.php), which provides stringent checks to facilitate automated harvesting using the same code used for validation at ingest.

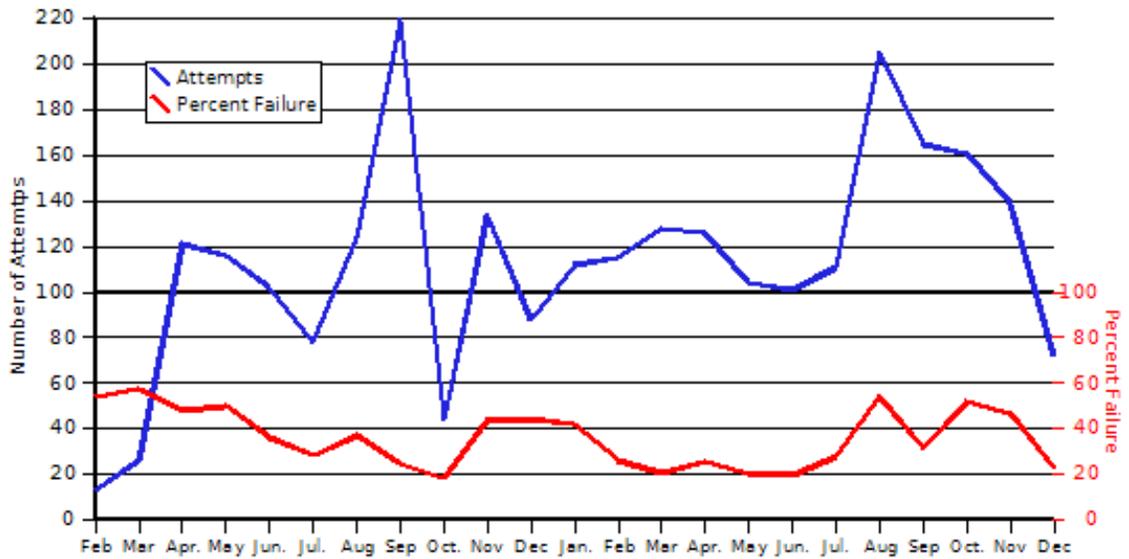

**Figure 2 - NSDL Harvest Failure Rate 2004-2005**

The notion of *incremental harvesting* is a fundamental part of the OAI-PMH model. Theoretically, a data aggregator should only need to do one initial full harvest, followed by repeated harvests that include modifications, deletions, and additions to the metadata from the data provider. In practice, incremental harvesting is often not possible due to two main problems:

- First, support for "deleted" records is inconsistent. As documented in the OAI Registry at UIUC[12], less than 50% of OAI-PMH data providers purport to persist ("forever", as defined in the OAI-PMH specification) deleted records. We have found, moreover, that some data providers that claim "persist" actually have less stringent perspectives on persistence and that a complete harvest is often the only reliable way to get an accurate snapshot of a data provider.

- Second, when OAI servers fail on any record during an incremental harvest, the start date cannot be updated. Similarly, any server instability can cause problems in determining an appropriate start date. The result is that a full harvest needs to be performed to "re-sync" the repository's view with that of the data provider.

As a result of these problems, initial harvest setup and regular harvests require constant monitoring. Emailed harvest results are sent to the CI harvest production team, who interpret them and contact the providers as necessary to correct OAI server protocol, XML, schema, and other errors. Weekly production meetings of the ingest team, together with a careful process of tracking harvest results and provider email exchanges, keep things relatively smooth, but the ongoing people cost is significant despite all efforts to automate.

### 4.3 Harvesting statistics

NSDL routinely harvests metadata from 113 collections via OAI. The harvesting discovers an average of 9250 items per collection. Each collection is re-harvested on an interval of between 1 to 3 months depending on the needs of the collection. Over the past two years NSDL has made over 2,600 harvest attempts.

We should note that not all collections run their own OAI service. Of the 114 collections, 37 are harvested from 8 OAI servers. This has resulted in economies for some collections. Additionally, many of the servers are based on shared code such as OCLC's OAICat[13] or Scout Portal Toolkit.

Our overall harvest success rate for the past two years is 64%. On a monthly basis our harvest failure rate has stubbornly hovered between 25-50%. This is illustrated in Figure 2. Periodically, major efforts have been made to reduce these failures (Sept 2004, Aug. 2005). While these efforts have pulled a great deal of new content into the repository, they did not succeed in lowering the failure rate over the long term.

The reasons for individual failures vary and, additionally, vary over time. Harvest failure categories are illustrated in Figure 3. Summarizing, our actual experience has shown that failures present themselves in equal measure within 3 broad categories: (i) a communications or system failure either at their server or with our OAI harvester, (ii) OAI protocol violations, and (iii) invalid XML data, XML schema non-compliance, or XML, URL, or UTF-8 character encoding. In fact, many of the OAI protocol violations are the result of these sorts of format errors, which result in an inability to process or even complete the OAI harvest.

---

[12] http://gita.grainger.uiuc.edu/registry/

[13] http://www.oclc.org/research/software/oai/cat.htm

Even the best maintained collections have difficulties at one point or another in their life cycle. Network and host availability issues are expected to impact harvesting, yet only 23% of harvest failures are due to such transient failures. Most harvest failures are due to data and protocol problems which require intervention by the metadata providers. In these cases, the specific causes of the failure must be thoroughly diagnosed, and the provider personnel contacted with actions needed to bring their OAI metadata back into compliance. Generally, harvesting cannot resume until the provider rectifies the problem(s). Note that CI staff attempt to find all co-occurring errors before contacting providers. While this is time consuming up front, it prevents repeated dialog on related errors.

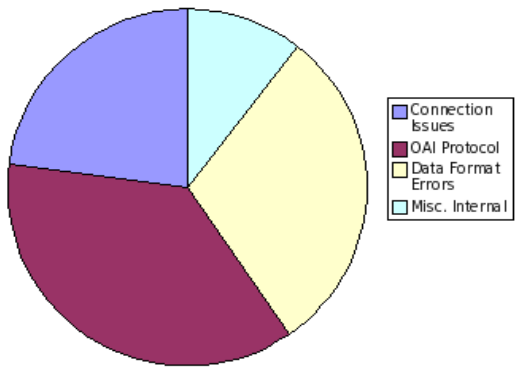

**Figure 3 - Harvest Failure Categories**

Email is the primary method of interaction with all providers. While some of this email is essentially templated machine-interpretable communication, much of it is human correspondence. A cross section of email archives of 8 representative providers revealed over 2,700 messages, or around 170 messages per provider per year.

The subjects of these emails are indicative of the difficulties in automating ingests from these feeds. 25% of all messages analyzed were automated reports of harvest failures. 39% were human or diagnostic messages, usually in response to failures. The remaining 36% were routine messages of successful harvests. The difficulty of setting up an OAI server and establishing harvesting is also apparent. On average, 98 messages were transmitted before NSDL was able to successfully retrieve its first harvest of a collection. In some cases, there were hundreds of email messages exchanged before a successful harvest occurred. Each of these messages corresponds to considerable human effort to resolve the problem.

## 5. INGEST PROCESSING

The goal of ingest processing is to transform the raw OAI feeds from metadata providers into metadata ready for storage in the Oracle-based metadata repository. These transforms address some metadata quality issues. Following the transforms, the metadata is staged in an XML file we call *dbInsert*. This is a list of metadata records similar to an OAI-PMH ListRecords response. The major difference is that a single metadata "record" in the dbInsert file contains both the originally harvested record (which remains unchanged) and the newly created normalized nsdl_dc generated by the transform process.

Whereas the plan has always been to ingest multiple rich metadata formats[14], the harvest-ingest process currently processes only two formats:

- *oai_dc:* the required OAI-PMH schema for unqualified Dublin Core.[15]
- *nsdl_dc:* an NSDL-specific application profile for qualified Dublin Core that includes extensions relevant for educational materials, as recommended by the DC Education Working Group [22].

There are two reasons for the delay in ingesting additional richer metadata formats:

First, as mentioned already, we have experienced considerable difficulty working with collections as they implement the minimal harvesting scenario: running an OAI-PMH server that provides the required oai_dc format. Many collections simply do not have the resources to take the next step and provide richer metadata after that initial implementation. nsdl_dc is considerably more expressive than oai_dc, yet only 50% of the collections provide metadata in that format. Only about 10% provide metadata in any of the other NSDL-supported metadata formats, providing little justification for CI to expend the effort necessary to process these formats.

Second, metadata quality, even with this minimal format, remains vexing. Our experience in improving the quality of DC records has been mixed. As described in [17], an initial approach involved "collection-specific" transforms, whereby we processed and corrected metadata on a collection-by-collection basis. We found, however, that in practice there was little consistency to the types of problems that arose within an individual collection's metadata, and "collection-specific" often evolved to "harvest-specific" corrections. Clearly this was not scalable.

We therefore evolved a more scalable strategy known as "safe transforms"[16], a process that takes oai_dc or nsdl_dc as input, fixes some common errors, applies some simple refinement techniques, and generates nsdl_dc as output. These transforms include:

- removing metadata fields with no information value (e.g., "no abstract submitted"),
- removing extraneous white-space,
- removing duplicate elements,
- qualifying easily recognized encoding schemes (e.g., URIs, well-known DCMI types, normalizing dc:language values), and

---

[14] http://metamanagement.comm.nsdl.org/IntroPage.html#standards

[15] http://www.openarchives.org/OAI/openarchivesprotocol.html#dublincore

[16] http://metamanagement.comm.nsdlib.org/safeXform.html

- correctly specifying and encoding URIs.

The last transform deserves additional explanation. As we describe in section 7, we need to "harden" the link between the metadata and the actual resource, so we can use that link for additional indexing. Reliable metadata->resource links are also useful for a number of other services. Thus, the MR ingest process "smartens up" dc:identifier fields: those that start *http://* or *ftp://* are designated with the DC URI encoding scheme if they can be automatically scrubbed into fetchable URLs. The ingest process also does an XML schema validation (via Xerces) and some additional validation on dc:identifier.dct:URI fields provided in Qualified Dublin Core by a collection. Those that fail these validation steps are downgraded to plain dc:identifier fields.

Some of these transforms, such as URI corrections, apply across formats, but many are specific to a metadata format, such as specific DCMI encodings and types. As a result, the introduction of each new metadata format requires expensive analysis of common problems and potential fixes in order to extend the "safe transform" philosophy. The scalability of this is questionable.

In the end, all of these transforms don't enhance the richness of the information in the metadata. Minimally descriptive metadata, like Dublin Core, is still minimally descriptive even after multiple quality repairs. We suggest that the time spent on such format-specific transforms might be better spent on analysis of the resource itself – the source of all manner of rich information.

## 6. METADATA STORAGE AND OAI EXPOSURE

The MR is implemented as an Oracle database. We chose an 'enterprise level' data store to allow rapidly deploying a repository capable of handling a very large number of metadata records. Also, local expertise made Oracle an attractive choice.

As a part of the redundancy and backup plan, two file servers are used to house the database. Metadata is processed through these two servers running separate Oracle instances. The metadata is inserted in and the XML metadata records are generated on one system, and the ready-to-expose records and supporting index tables are transferred to a separate system that feeds the OAI-PMH service. This separation of ingest, XML generation, and exposure has allowed for flexibility in configuration and backup of the source data and the served XML records.

Along with various logging and administrative data tables, the MR database schema contains three sets of tables: a set of five tables for storing data as it is parsed on input into the system, a set of four tables that contain the generated XML formats that are used for OAI-PMH serving, and a set of seven tables that contain the combined OAI-formatted data and index tables optimized for retrieval by our java-based OAI server.

### 6.1 Data flow through the MR database

An initial (SAX) parse of the inbound dbInsert XML metadata records separates them into the two sets of records that the safe-transform process creates – the normalized nsdl_dc records created by the safe-transform, and the original records harvested from the OAI provider. These original records are then stored as a single string with their own originator date stamp and schema identification.

The nsdl_dc records generated by the normalization process are shredded into element-value pairs and stored. Element names and their source nsdl_dc schema and schemes are coded and identified in reference tables. This structure was chosen to facilitate analysis and modifications of specific elements within the normalized nsdl_dc records across all metadata records. It is also used to generate a rudimentary resource index by extracting all identifier elements of all metadata records that are URI-like.

As metadata records are inserted, the records for OAI exposure are also generated. NSDL currently produces five distinct OAI formats from this metadata:

- *nsdl_dc* is the normalized Qualified Dublin Core version of the harvested metadata records.
- *oai_dc* is the simple Dublin Core record required of any OAI-PMH data provider. This is a dumbed-down version of the normalized nsdl_dc.
- *nsdl_links* indicates relationships between metadata records. Currently the only relationship represented is collection membership – each record is a member of a collection, which is represented by a metadata record for the collection.
- *nsdl_search* is a combined format that includes the above three formats as well as the original 'native' harvested format. This format is not released to the public, as the provider of the native format may not wish to share their metadata, and it is currently used only for the NSDL's search-index process.
- *nsdl_all* is the same as nsdl_search except that the 'native' metadata record will not be present if the metadata provider has requested that their metadata not be made publicly available.

All of these five served formats are generated as large strings and stored in the staging tables on the ingest server. A timed process runs on the serving database that queries the staging database for new entries. New entries are gathered and the tables required to serve OAI-PMH are populated with the new or updated entries. The serving Oracle instance contains views and some level of de-normalization of table data in order to optimize the queries that the java OAI server uses to service requests.

### 6.2 Lessons from the MR implementation

Oracle has proven to be a flexible data storage tool, but the cost for configuration and operations has been high. Configuring and tuning the database to perform optimally has taken considerable time and effort, and the on-going management of the database has required more-than-expected personnel resources as well. The ingest-staging database contains about 55GB of data, and the current OAI serving database contains approximately 53GB of data.

Because OAI records have datestamps that are used in incremental harvests, it is crucial that the datestamp associated with an OAI record be calculated appropriately. This required some rather arcane processing. OAI-PMH idempotency requirements mandate that a request for records between the dates D and D+Δ will always return the same records, if they have not been updated in the meantime. Since the record datestamps must be generated significantly before we expose the OAI record in our tables, to meet the OAI-PMH idempotency requirements we must post-date all inbound metadata records by three hours. If we didn't postdate into the future, then harvest requests for very recent metadata could potentially be missing any OAI records that had not yet been generated for exposure.

Throughput for processing harvested records in the current production environment runs between 5000-10000 records per hour. This depends greatly on the transfer rate from the originating OAI server, the number of records to process, and the density of those records.

Early in the life of the repository, some errors in content propagated through to the data store. As our error detection and correction efforts have improved, most of these errors have been corrected, but some, particularly from collections that are no longer available, are still in the system. The people cost of correcting these errors is too high, so we continue to serve a small percentage of OAI records with XML schema errors.

Our current ingest process, fairly robust after two major rewrites and numerous bug fixes, is still vulnerable to occasional UTF8 encoding and XML Schema validation errors creeping into newly stored records. These errors often go unobserved for weeks until some downstream user or service stumbles on them.

### 6.3 Functionality of the RDBMS-based MR

Overall, the Oracle RDBMS has been successful as a tool for metadata storage, meeting the original requirements. However, as the NSDL has matured, the requirements have grown. We note two areas where the RDBMS has been problematic in extending the functionality of the NSDL.

We increasingly find that storing and querying an expanding set of relationships among library entities – resources, metadata, annotations, standards, and providers – is essential. Handling queries such as "find all the resources contributed by DLESE that meet the California middle school standard for earth science" is critical for building the types of customized applications of the NSDL that we envision. While the RDBMS design contains a "links" (relationships) table, it lacks the expressiveness of ontology-based relationships. Furthermore, composing transitive queries across entity-relationship graphs is cumbersome and may encounter expensive blow-ups in the number of joins.

The table design of the MR is based on the notion of structured data – metadata elements and values. However, following the initial release, CI has tried to move to less structured forms of data and, in fact, into the creation and storage of content itself – e.g., lesson plans, curricula, annotations, etc. The MR-based architecture, which imposes a strict bifurcation between "metadata" and "data", has interfered with the effort to create a unified data repository that can flexibly accommodate a range of structured, un-structured, and semi-structured data.

### 7. SEARCH

The NSDL Search Service is, essentially, the first customer of MR records exposed via OAI, and the information in the NSDL search index determines whether a resource can be discovered via searching at the main NSDL portal[17]. In fact, many collections check to see if their metadata has been integrated into the NSDL by doing "known item" searches at nsdl.org. Thus, the search service is sometimes used to discover ingest errors such as missing or incorrect metadata.

The current production search service is based on the metadata aggregation model, and it is sometimes referred to as "metadata-centric." As the limitations of this model have come to light, and additionally, as nsdl.org users have complained about finding duplicates in their search results, we have moved to create a "resource-centric" search service, both to avoid duplicates in search results and to position us to include richer information, such as context and less structured metadata, in determining nsdl.org search results.

### 7.1 Metadata-Centric Search

The metadata-centric search service starts with an OAI harvest from the MR. The XML metadata received is parsed, then indexed using Lucene, an open source search engine. The index contains the normalized nsdl_dc metadata, the "raw native" metadata, and some additional information, such as NSDL collection membership. Each metadata record becomes a document in the Lucene index – a document roughly equates to a "hit" in search results. Thus metadata-centric search results contain a "hit" for each relevant metadata record in the index. The indexed metadata is updated incrementally – only records modified since the previous harvest are requested from the MR OAI server, and the results are used to update the existing Lucene index.

We also fetch and index the textual content of resources described by the metadata. The search service looks in the dc:identifier.dct:URI fields exposed in the normalized nsdl_dc from the MR for URLs we can fetch. Then the search service uses Nutch, an open source web crawler, to fetch the content and manage it. (Currently the Nutch software comes with code to retrieve content via http and ftp). Nutch stores the URLs and the fetched content (both as received and as text ready to be indexed) for efficient storage and access and also provides a mechanism to refresh stale content automatically. As of January 25, 2006, the production search index contained 1,056,407 Lucene documents, representing all the "active" metadata records from the MR. (Note that this number does not reflect approximately 280,000 MR OAI metadata records marked deleted.) Approximately 7500 of these Lucene documents have no URL resource identifier, meaning there was no resource URL that passed our validation.

### 7.2 Resource-Centric Search

We have a number of reasons for moving from a data model that is metadata-centric to one that is resource-centric. For example, we receive metadata records from a large number of providers, and some of those are about the same resource. Rather than a simple metadata repository that stores these as separate records, they should be related to a common resource "entity", which is currently not represented in the MR data model. In the future, we also want to express the relationships between resources and other information, such as annotations and standards alignments. Finally, we wish to inter-relate resources themselves, such as their co-existence within a lesson plan or curriculum. That resource-centric model is the subject of current work on an NSDL Data Repository (NDR), which will replace the MR [19].

---

[17] http://nsdl.org

Independent of that work, we have been transitioning to a more resource-centric search service, currently using the metadata repository, but later the NDR. Whereas the current search engine has a one-to-one mapping from metadata record to "hit", this work will map hits to resources – independent of the number of metadata records about that resource.

In order to do this, we need to infer resource equivalence from the MR, which sits at the end of a data flow that up to this point is entirely metadata focused. We determine equivalence by exploiting the identifiers in the nsdl_dc records that we harvest from the MR.

We should note, however, that the URL in an item record does not automatically correspond to an actual link to the real digital resource described by the metadata. We have found that some metadata providers shortcut the effort to actually insert a unique item URL in the DC record by using the same collection "splash page" URL for a set of item records. This indicates that the methods we describe below for determining resource equivalence need to also account for "fuzzy equivalence" between metadata records – i.e., whether two records that purport to describe the same resource (measured by URL equivalence) are really "about" the same content [16].

At this point, however, we are taking two approaches to determining equivalence using the URLs in the metadata records.

### 7.2.1 Resource equivalence phase I: URL normalization

The URI specification [5] enumerates steps to normalize URLs to determine if they are equivalent. This includes ensuring the scheme and hostname are lower case, the default port is not specified, an empty absolute path is represented as a trailing slash, and so on. The search service addresses most of this URL normalization with java.net.URI methods; the remaining pieces are addressed with additional java code.

Initially we took a naïve approach that created a "resource" (a Lucene document) for each dc:identifier and dc:identifier.dct:URI in the OAI metadata. However, this naïve approach had the undesirable effect of increasing the number of documents in the Lucene index by almost 50%: at that time we had slightly more than 1 million documents in the metadata-centric index, and almost 1,500,000 documents in this naïve resource-centric index. In examining the causes, we learned that there are approximately 1,500,000 dc:identifier fields (in various flavors) but the number of fetchable URLs is closer to 1 million.

Before choosing a different algorithm and making a similar mistake, we chose to examine our dc:identifier fields and our metadata records more carefully. This involved writing some tools to examine Lucene index contents, as well as performing SQL queries against our Oracle database. Because of our decision to split our normalized nsdl_dc into elements in the Oracle DB, getting information such as "what do records with multiple fetchable resource identifiers look like?" and "how many metadata records have 2 or more fetchable resource identifiers" has been difficult and is still in progress. As of January 26, 2006, we count approximately 180,000 metadata records with 2 or more fetchable resource identifiers.

### 7.2.2 Resource equivalence phase II: comparison of fetched content via MD5Hash

The Nutch application creates an MD5Hash for fetched content to facilitate comparison. In our current work, we will use these checksums to determine if fetched content is equivalent. If so, the normalized resource URLs (and matching NDR resource digital objects) will be marked as part of an equivalence class, and the corresponding Lucene documents in the search index will be merged into a single Lucene document for the resource. This phase has not yet been implemented, but we will report results in a future paper.

## 8. CONCLUSION

Over the last three years the NSDL CI team has learned that a seemingly modest architecture based on metadata harvesting is surprisingly difficult to manage in a large-scale implementation. The administrative difficulties result from a combination of provider difficulties with OAI-PMH and Dublin Core, the complexities in consistent handling of multiple metadata feeds over a large number of iterations, and the limitations of metadata quality remediation.

More problematic are the shortcomings of the architecture as the basis for a service-rich digital library. As noted in previous sections, the centrality of structured metadata interferes with the intermingling of potentially more valuable unstructured and structured data and the rich relationships among these data entities. Even the implementation of search, a basic digital library service, is hampered by the need to recover a resource-centric view from a dataflow that is solely metadata-centric.

Arguably, it makes more sense to create an architecture that begins with a resource-centric view (e.g., a set of resource URIs from a web crawl) and carries that view through the entire model. The CI team is now implementing such an architecture, based on the notion of an information network overlay. This architecture emphasizes the integration of multiple information entities and their rich relationships, while focusing on creating and expressing context for resources. The implementation of this architecture has confronted a number of hurdles, due to its reliance on cutting edge technologies. But we expect release during 2006, and it will lead to an NSDL that provides a uniquely rich, flexible, and collaborative digital library environment focused on STEM education.

## ACKNOWLEDGMENTS

The work described here is based upon work supported by the National Science Foundation under Grants No. 0227648, 0227656, and 0227888. Any opinions, findings, and conclusions or recommendations expressed in this material are those of the author(s) and do not necessarily reflect the views of the National Science Foundation. The authors acknowledge the efforts and support of the entire CI team. Bill Arms, Diane Hillmann, and Jon Phipps deserve special recognition due to their fundamental role in the design and implementation of the system described here.